\documentclass[aps,12pt,showpacs]{revtex4}

\usepackage{epsfig}

\begin{document}
\rightline{HD-THEP-04-26}
%\title{Energy density in an expanding universe is correctly seen by
%        Unruh's Detector}
\title{\Large Energy Density in Expanding Universes 
as Seen by Unruh's Detector}

\author{\Large Bj\"orn Garbrecht$^*$, Tomislav Prokopec}

\email[]{B.Garbrecht@ThPhys.Uni-Heidelberg.De}
\email[]{T.Prokopec@ThPhys.Uni-Heidelberg.De}

\bigskip

\affiliation{Institut f\"ur Theoretische Physik, Universit\"at Heidelberg,
             Philosophenweg 16, D-69120 Heidelberg, Germany}

\date{\today}

\begin{abstract}
We consider the response of an Unruh detector to scalar fields in an
expanding space-time. When combining transition elements of the scalar field
Hamiltonian with the interaction operator of detector and field, one
finds at second order in time-dependent perturbation theory a
transition amplitude, which actually dominates 
 in the ultraviolet over the first order contribution.
In particular, the detector response faithfully reproduces
the particle number implied by the stress-energy of a minimally coupled
scalar field,
which is inversely proportional to the energy of a scalar mode. This
finding disagrees with the contention that in de Sitter space, the
response of the detector drops exponentially with particle energy and therefore
indicates a thermal spectrum.
\end{abstract}


\maketitle

\section{Introduction}
There seems to be a striking bias between the particle number density as
inferred from the stress-energy tensor of a scalar field in an expanding
Universe and the response of an Unruh detector.
For particle energies  $\omega$ larger than 
the Hubble rate $H$, the first quantity falls as $1/\omega^2$, whereas the
latter drops in de Sitter space
exponentially, like 
${\rm e}^{-2\pi \omega/H}$~\cite{GibbonsHawking:1977,Higuchi:1986,BoussoMaloneyStrominger:2001,Lapedes:1978,GarbrechtProkopec:2004},
which has led to the notion that this space-time is endowed with a temperature.

A closely related prediction of quantum theory in curved
space-time~\cite{BirrellDavies:1982} is
the possible generation of the observed primordial density fluctuations 
in the expanding Universe during an inflationary
epoch~\cite{MukhanovChibisov:1981,GuthPi:1982,Hawking:1982,Starobinsky:1982},
which can be calculated by finding a Bogolyubov rotation
between the incoming and the outgoing vacuum, such that
the free Hamiltonian is diagonal in terms 
of creation and annihilation operators~\cite{Parker:1969}.
The particle number found from the transformed
creation and annihilation operators is then also
proportional to $1/\omega^2$, in disagreement with the detector
response at first order in perturbation theory, which is due to the fact that
the incoming vacuum state is of purely negative frequency,
while this does not hold for the outgoing vacuum.

In fact, the problem does not occur for the celebrated Hawking
effect~\cite{Hawking:1974},
the emission of radiation with thermal spectrum by a black hole.
For an observer far from a black hole, the background space is asymptotically 
flat, and
the canonical Hamiltonian agrees with the stress-energy tensor.
The Hamiltonian is diagonal in the basis where a positive frequency
mode corresponds to a particle and negative frequency to an antiparticle.
The particle number can be observed by an Unruh detector~\cite{Unruh:1976},
which 
interacts under these circumstances with the scalar field through the
familiar processes of absorption, spontaneous and stimulated emission, 
calculated at first order in time-dependent perturbation theory, which
we review in section~\ref{section:UnruhFlat}. It
hence counts the number of positive frequency modes as particles,
and there is no conflict with the result obtained from the Bogolyubov
transformation. 

In order to resolve the disagreement for the expanding Universe case,
it is pointed out in
Refs.~\cite{Lapedes:1978,BoussoMaloneyStrominger:2001} that, instead of
diagonalizing the Hamiltonian, one can obtain a Bogolyubov transformation
related to a thermal spectrum by matching the modes of the scalar field 
to a certain basis defined in static coordinates. Yet, this approach does
not resolve the discrepancy between the first order detector response
and the component of the stress-energy tensor describing the energy density.

In section~\ref{section:UnruhExpanding}, we suggest how to resolve this seeming
contradiction. We point out that in expanding space-times, it is
not suitable to describe scalar particle creation and detection by a mixing
of positive and negative frequency modes. The main source for the
detector response in the ultraviolet domain
is rather given by the pair creation amplitude of the scalar
Hamiltonian. When combining this at second order in time-dependent
perturbation theory with the transition elements of the detector
and field interaction Hamiltonian, 
we find a response in accordance with the
energy density of a minimally coupled scalar field.

\section{Stress-Energy}\label{section:StressEnergy}
This section contains a review of adiabatic expansion of
the stress-energy tensor of a scalar field in a
Friedmann-Lema\^\i tre-Robertson-Walker~(FLRW)
Universe~\cite{ParkerFulling:1974,HabibMolina-ParisMottola:1999}
and also serves us to introduce some notations.

We employ conformal coordinates, in which the metric has the form
\begin{equation}
g_{\mu\nu}=a^2(\eta){\rm diag}(1,-1,-1,-1),
\end{equation}
and denote the derivative %\emph{w.r.t.} 
with respect to 
conformal time $\eta$ as
$'\equiv d/d\eta$. The scalar field Lagrangean is
\begin{equation}
\sqrt{-g}\mathcal{L}
  =\sqrt{-g}\left(\frac{1}{2}g_{\mu\nu}\partial^\mu\phi\partial^\nu\phi
-\frac{1}{2}m^2\phi^2-\frac{1}{2}\xi R\phi^2\right),
\end{equation}
where $R$ denotes the curvature scalar, 
$g = {\rm det}(g_{\mu\nu}) = a^8$, and the action is
\begin{equation}
S=\int d^4x \sqrt{-g}\mathcal{L}.
\end{equation}
From the variation
\begin{equation}
\delta S=-\frac{1}{2}\int d^4x \sqrt{-g} T^{\mu\nu}\delta g_{\mu\nu},
\end{equation}
we find the stress-energy tensor to be
\begin{eqnarray}
T^{\mu\nu}&=&\partial^\mu\phi\partial^\nu\phi
-\frac{1}{2}g^{\mu\nu}\left(
g_{\varrho\sigma}\partial^\varrho\phi\partial^\sigma\phi-m^2\phi^2\right)
%\nonumber\\
+\xi G^{\mu\nu}\phi^2
+\xi\left(g^{\mu\nu}g_{\varrho\sigma}\nabla^\varrho\nabla^\sigma
-\nabla^\mu\nabla^\nu\right)\phi^2
\label{stress-energy},
\end{eqnarray}
where $G^{\mu\nu}=\frac{1}{2}Rg^{\mu\nu}-R^{\mu\nu}$ is the Einstein tensor,
$R^{\mu\nu}$ the Ricci tensor and $\nabla$ denotes covariant derivative.

The field $\phi$ obeys the Euler-Lagrange equations of motion, 
which take the form
\begin{equation}
%g^{\mu\nu}\partial_\mu\partial_\nu\phi
%-g^{\mu\nu}\Gamma^\nu_{\mu\kappa}\partial^\kappa\phi
\left[\nabla^2+m^2+\xi R\right] \phi(x)=0.
\end{equation}
In a FLRW-background, they become
\begin{equation}
\left[\partial_\eta^2+2\frac{a'}{a}\partial_\eta
 + \left(-\nabla^2+a^2m^2\right)
 + 6\xi \frac{a''}{a} \right]\phi(x)
 = 0
\,.
\end{equation}
Upon the substitutition $\varphi=a\phi$, 
the damping term $\propto\partial_\eta \phi$ drops out, and
the following single mode decomposition of $\varphi$
then generally holds: 
\begin{equation}
  \varphi(x) = \int \frac{d^3 k}{(2\pi)^3}
              \Big(
                   {\rm e}^{i\mathbf{k}\cdot\mathbf{x}}
                    \varphi(\mathbf{k},\eta)
                            a({\mathbf{k}})
                +  {\rm e}^{-i\mathbf{k}\cdot\mathbf{x}}
                     \varphi^*(\mathbf{k},\eta)
                            a^\dagger({\mathbf{k}})
              \Big)
\,.
\label{varphi}
\end{equation}
Here $a({\mathbf{k}})$ and $a^\dagger({\mathbf{k}})$ denote the 
annihilation and creation operators for the mode with 
a comoving momentum $\mathbf{k}$ and are defined by 
$a^\dagger({\mathbf{k}})|0\rangle = |\mathbf{k}\rangle$,
$a({\mathbf{k}})|\mathbf{k}^\prime\rangle 
  = (2\pi)^3\delta^3(\mathbf{k}-\mathbf{k}^\prime)|0\rangle$,
where $|0\rangle$ denotes the vacuum state
and $|\mathbf{k}\rangle$ the one-particle state with 
momentum $\mathbf{k}$.
The mode functions $\varphi(\mathbf{k},\eta)$ satisfy the following 
equation:
\begin{equation}
\Big(\partial_\eta^2+\big(\mathbf{k}^2+a^2m^2\big)
   + (6\xi-1)\frac{a''}{a}
 \Big)
\varphi(\mathbf{k},\eta)=0
 \label{phieq}
\,.
\end{equation}

Upon inserting~(\ref{varphi}) into~(\ref{stress-energy})
and taking expectation value with respect to the vacuum $|0\rangle$,
%we get for the components of the stress-energy tensor modes
we get for the expectation values of the components of 
the stress-energy tensor
%$T^{\mu\nu}(\mathbf{k},\eta)$
% $\big(\langle0|T^{\mu\nu}(x)|0\rangle = \int [{d^3k}/{(2\pi)^3}]
%                    T^{\mu\nu}(\mathbf{k},\eta) \big)$
%
\begin{eqnarray}
\langle0|T^{00}(x)|0\rangle &\!=\!& 
 \int \frac{d^3k}{(2\pi)^3}
 \frac{1}{a^{6}}\Bigg\{\!
 \left[\omega^2+
 \frac{1-6\xi}{2}\left(\frac{a'^2}{a^2}-\frac{a''}{a}\right)\right]|\varphi|^2
%\nonumber\\
-\frac{1-6\xi}{2}\frac{a'}{a}\partial_\eta|\varphi|^2
+\frac{1}{4}\partial_\eta^2|\varphi|^2\label{ergden}
\Bigg\}
\,,
 \quad
 \label{enden}
\\
\langle0|T^{0i}(x)|0\rangle &\!=\!& 
\langle0|T^{i0}(x)|0\rangle =
 \int \frac{d^3k}{(2\pi)^3}
         \frac{k^i}{a^6}\Im(\varphi^\prime\varphi^*)
\,,
 \label{stress}
\\
\langle0|T^{ij}(x)|0\rangle &\!=\!& 
%T^{ij}(\mathbf k,\eta) &\!=\!& 
 \int \frac{d^3k}{(2\pi)^3}
 \frac{1}{a^{6}}\Bigg\{\!
\left[k^ik^j+\delta^{ij}
\frac{1\!-\!6\xi}{2}
 \left(\frac{a'^2}{a^2}-\frac{a''}{a}\right)\right]|\varphi|^2
\nonumber\\
&&\hskip 2cm 
 - \delta^{ij}\frac{1\!-\!6\xi}{2}\frac{a'}{a}\partial_\eta|\varphi|^2
 +\delta^{ij}\left(\frac{1}{4}\!-\!\xi\right)\partial_\eta^2|\varphi|^2
\Bigg\}
\,,\qquad
 \label{momden}
\end{eqnarray}
where we defined $\omega = \sqrt{\mathbf{k}^2+a^2m^2}$.
Note that, as a consequence of the isotropy of FLRW space-times,
 $\langle0|T^{\mu \nu}(x)|0\rangle$ 
%and $\langle0|T^{ij}(x)|0\rangle$ $(i\neq j)$
$(\mu\neq \nu)$ vanishes. 
Indeed, the space-time isotropy implies 
that $\varphi$ is a function of the momentum magnitude
$k\equiv |\mathbf{k}|$ and $\eta$ only, such that, 
when the contributions to the stress-energy tensor of opposite momenta
are added, a cancellation occurs.

An analytic solution to Eq.~(\ref{phieq}) can only be found for special
$a(\eta)$. We therefore adapt the approach of adiabatic expansion and
start with the WKB {\it ansatz}
\begin{equation}
\varphi(\mathbf k,\eta)=
\alpha(\mathbf k)\big(2W(\mathbf k,\eta)\big)^{-\frac 12}\,
{\rm e}^{-i\int\limits^{\eta} d \eta' W(\mathbf k,\eta')}
+ \beta(\mathbf k)\big(2W(\mathbf k,\eta)\big)^{-\frac 12}\,
{\rm e}^{i\int\limits^{\eta} d \eta' W(\mathbf k,\eta')}
\label{WKB},
\end{equation}
with the normalization condition $|\alpha|^2-|\beta|^2=1$.
Then, we find from~(\ref{phieq})
\begin{equation}
W^2=\omega^2-(1-6\xi)\frac{a''}{a}+\frac{3}{4}\frac{W'^2}{W^2}
-\frac{1}{2}\frac{W''}{W}
%\qquad 
%  \omega^2(\mathbf k,\eta)=\mathbf{k}^2+a(\eta)^2m^2
\,.
\end{equation}
We take for the vacuum $|0\rangle$ the purely negative frequency state
at infinitely early times, that is $\alpha=1$ and $\beta=0$.
This choice is well motivated by cosmological inflationary models, 
and it is a standard choice for studies of de Sitter 
space~\cite{GibbonsHawking:1977,Higuchi:1986,BoussoMaloneyStrominger:2001,Lapedes:1978,GarbrechtProkopec:2004,ChernikovTagirov:1968,BunchDavies:1978}, 
as well as for general FLRW
space-times~\cite{ParkerFulling:1974,HabibMolina-ParisMottola:1999}.
Up to second adiabatic order,
that is to second order in derivatives \emph{w.r.t.} $\eta$, the
stress-energy~(\ref{enden}--\ref{momden}) is
\begin{eqnarray}
\langle0|T^{00}_{(2)}(x)|0\rangle \!&=&\!
 \int \frac{d^3k}{(2\pi)^3}
 \frac{1}{a^{6}}
         \Bigg\{
                \frac{\omega}{2}+\frac{1-6\xi}{4\omega}\frac{a'^2}{a^2}
             +  \frac{1-6\xi}{4}\frac{a'^2}{a^2}\frac{a^2m^2}{\omega^3}
             +  \frac{1}{16}\frac{a'^2}{a^2}\frac{a^4m^4}{\omega^5}
         \Bigg\}
\,,
\label{T:adiabatic}
\\
\langle0|T^{0i}(x)|0\rangle &\!=\!& 
\langle0|T^{i0}(x)|0\rangle = 0
\,,
\\
\langle0|T^{ij}_{(2)}(x)|0\rangle &\!=\!& 
 \int \frac{d^3k}{(2\pi)^3}
\frac{\delta^{ij}}{a^{6}}\Bigg\{
               \frac{\omega}{6}
              -\frac{a^2 m^2}{6\omega}
+\frac{1-6\xi}{12\omega}\left[3\frac{a'^2}{a^2}-2\frac{a''}{a}\right]
%\nonumber\\
\!\!+\!\!\frac{1-6\xi}{6}
\left[\frac{a'^2}{a^2}-\frac{a''}{a}\right]\frac{a^2 m^2}{\omega^3}
\nonumber
\\
&&\hskip 2.cm
 +\,
\left[\left(\frac{11}{48}-\frac{3}{2}\xi\right)\frac{a'^2}{a^2}
 -\frac{1}{24}\frac{a''}{a}\right]\frac{a^4 m^4}{\omega^5}
+\frac{5}{48}\frac{a'^2}{a^2}\frac{a^6 m^6}{\omega^7}
\Bigg\}
\,,
\label{momden2}
\end{eqnarray}
where we made use of $\int d^3k k^i k^j = (\delta^{ij}/3)\int d^3 k k^2
                       = (\delta^{ij}/3)\int d^3 k (\omega^2 - a^2 m^2)$
and $\Im\big(\varphi^\prime\varphi^*)\big) 
     = - 1/2$.
For both, the zeroth and second adiabatic order contributions
seperately, the covariant conservation
$\nabla_\mu T^{\mu 0}_{(2)}=\partial_\eta T^{00}_{(2)}
+5({a'}/{a})T^{00}_{(2)}
+3({a'}/{a})T^{ii}_{(2)}=0$ holds. When integrated over momentum space,
the zeroth order term yields a quartic divergence, corresponding to the
vacuum energy, also referred to as cosmological term, and the second order 
contribution diverges quadratically.
In order to deal with the infinity at second adiabatic order, it is suggested
either to rescale Newton's constant $G$ by an infinite
amount~\cite{BirrellDavies:1982,ParkerFulling:1974,Christensen:1976},
or to simply discard this term, which then does
not contribute to the observable particle
number~\cite{HabibMolina-ParisMottola:1999}.

Since an ultraviolet regularization should not affect the infrared domain,
the latter procedure is in conflict with
the amplification of quantum fluctuations at horizon crossing during
inflation~\cite{MukhanovChibisov:1981,GuthPi:1982,Hawking:1982,Starobinsky:1982},
leading to the observed primordial density fluctuations, which is a sound
prediction of quantum theory in curved space-times.
In that calculation, a subtraction of
the second adiabatic order terms is not performed,
and would make no sense.

The other alternative, the renormalization of $G$,
seems to disagree with perturbative quantum gravity.
Since gravity is a perturbatively nonrenormalizable theory,
order by order in loops,
new coupling constants for higher order geometric invariant terms are 
renormalized~\cite{'tHooftVeltman:1974,Donoghue:1994}, but not the leading
Newton's constant $G$. In fact, perturbative quantum gravity reproduces at
tree level known
classical metrics of general relativity and predicts quantum corrections at
loop
order~\cite{Donoghue:1994,Bjerrum-Bohr:2002ks,Bjerrum-Bohr:2002kt,DonoghueHolsteinGarbrechtKonstandin:2001,Bjerrum-Bohr:2002sx}
without shifting $G$ by any amount.
It is however interesing to note that the adiabatic regularization approach
can be useful in identifying divergences occurring at loop order in
quantum electrodynamics. In
Refs.~\cite{Kluger:1991ib,Kluger:1992gb} it is shown that, for pair
creation in a stationary electric field, one recovers the familiar
logarithmic divergence of the photon vacuum polarization.

In the following, we take the point of view that the second
adiabatic order contributions are observable energy and momentum
densities, and they should not be removed but need to be regulated by a
cutoff. In fact, in section~\ref{section:UnruhExpanding} we point out
that an Unruh detector observes an amount of particles which accords
with the energy density at second adiabatic order. 
The zeroth order term corresponds to the vacuum energy
and is subtracted. We shall come back to this issue in the discussion
section~\ref{section:Conclusions}.

\section{Detector in Flat Space-Time}\label{section:UnruhFlat}
We consider Unruh's detector~\cite{Unruh:1976}, a heavy particle moving along
a trajectory $x=x(t)$, where $t$ is its proper time. The Hamiltonian of
the detector is given by $H=H_0+\delta H$, where $H_0$ is the
unperturbed (time independent)
 Hamiltonian and $\delta H$ accounts for the interaction with
the scalar bath $\phi$, which we assume to be in a state $|i\rangle$.
While we treat the detector by the
means of nonrelativistic quantum mechanics, the nature of $\phi$
as a quantum field is of importance.
The situation is therefore very similar to absorption and emission of photons
by an atom, a discussion of which can be found
in any textbook on quantum mechanics.

Let us first define the set of unperturbed eigenstates of the detector
by
\begin{equation}
|m^0,t\rangle={\rm e}^{-iH_0 t}|m^0\rangle={\rm e}^{-iE_m t}|m^0\rangle,
\end{equation}
which acquire nondiagonal transition amplitudes through the
interaction Hamiltonian
\begin{equation}
\delta H=\hat{h}\phi(x).
\label{hamilton:interaction}
\end{equation}
The operator $\hat{h}$ is a quantum mechanical operator determined by the
inner structure of the detector and has the elements
$h_{mn}=\langle E_n|\hat{h}|E_m\rangle$, while $\phi$ is a quantum field
operator to be expanded in modes suitable for the given background spacetime,
which is flat space in this section.

Initially, at time $t_0$, the detector is in the state $|m^0,t_0\rangle$,
which evolves under the action of the full Hamiltonian $H$ into $|m,t\rangle$
at some time $t$. We want to calculate the amplitude of exciting the detector
from $E_m\rightarrow E_n$, hence
\begin{equation}
\mathcal{M}_{mn}=
  \langle n^0,t|m,t\rangle =\langle n^0|{\rm e}^{iH_0 t}|m,t\rangle
=\langle n^0|m,t\rangle_I,
\end{equation}
where the interaction state is given to first order in the von 
Neumann series as
\begin{equation}
|m,t\rangle_I 
=|m^0\rangle-i\int_{t_0}^t dt' 
         {\rm e}^{iH_0 t'}\delta H(t'){\rm e}^{-iH_0 t'}|m^0\rangle
\,,
\end{equation}
and $|n^0\rangle \equiv |n^0,t=0\rangle$,
$|m^0\rangle \equiv |m^0,t=0\rangle$.
We find
\begin{equation}
\mathcal{M}_{mn}=\delta_{mn}-i\int_{t_0}^t dt'
{\rm e}^{i(E_n-E_m)t'}\langle n^0|\delta H(t')|m^0\rangle
\,,
\label{amplitude:1st order}
\end{equation}
which, upon inserting~(\ref{hamilton:interaction}), reads
\begin{equation}
\mathcal{M}_{mn}=\sum\limits_f \langle f|\delta_{mn}
-i\int_{t_0}^t dt' 
    {\rm e}^{i(E_n-E_m)t'} h_{mn} \phi\left(x(t')\right)|i\rangle
\,.
\end{equation}
Here the scalar field has undergone a transition from
$|0\rangle$ to some element $|f\rangle$
of an orthonormal set of final states, which we summed over.
The probability of a transition from $E_m\rightarrow E_n$, where $n\not= m$
is hence
\begin{equation}
\mathcal{P}_{mn}=|\mathcal{M}_{mn}|^2=\sum\limits_f \left|\langle f|
\int_{t_0}^t dt' {\rm e}^{i(E_n-E_m)t'} h_{mn} \phi\left(x(t')\right)
|i\rangle\right|^2
.
\label{P:correct}
\end{equation}
We sum over the basis $f$ and obtain
\begin{equation}
\mathcal{P}_{mn}=|h_{mn}|^2 \int_{t_0}^t dt' \int_{t_0}^t dt''
{\rm e}^{i(E_n-E_m)(t'-t'')} \langle i|\phi\left(x(t'')\right)
\phi\left(x(t')\right)|i\rangle.
\end{equation}
Setting $t_0=0$, $\Delta E=E_m-E_n$ , and taking the derivative \emph{w.r.t.} $t$
gives the following result for the response function
${\mathcal{F}}_{\rm flat}\equiv
      \mathcal{P}_{mn}/|h_{mn}|^2$~\cite{GarbrechtProkopec:2004},
\begin{equation}
\frac{d{\mathcal{F}}_{\rm flat}(\Delta E)}{dt}
   =\int_{-t}^t d{\Delta t} {\rm e}^{i\Delta E \Delta t}
      \langle i|
                \phi\left(x(-\Delta t/2)\right)\phi\left(x(\Delta t/2)\right)
       |i\rangle
\,,
\label{response function}
\end{equation}
where we have assumed that the state of the scalar field respects 
the time translation invariance of flat space.
%(Note that ${\cal P}_{nn}$ from the `elastic' transitions
%$E_n\rightarrow E_n$ 
%are time independent, and hence do not contribute to $d{\cal F}/dt$.)

Now consider Minkowski space filled with particles of the spectrum
$|i\rangle\equiv \prod_{\mathbf{k}}\otimes|\nu(|\mathbf{k}|)\rangle$ 
and mass $m$, where
$\nu(|\mathbf{k}|)$ denotes the particle number per mode,
which we assume to be isotropic. We expand the field operator as in
Eq.~(\ref{varphi}), and take $a\equiv 1$, $\eta\equiv t$ in flat space-time.
%({\it cf.} Eq.~(\ref{varphi}))
%\begin{equation}
%\phi(x)\equiv\frac{\varphi(x)}{a}=\frac{1}{a}\int\frac{d^3k}{(2\pi)^3}
%  {\rm e}^{i\mathbf{k}\cdot\mathbf{x}}
%\left(\varphi(\mathbf{k},t)a(\mathbf{k})+
%\varphi^*({-\mathbf{k}},t)a^\dagger(-\mathbf{k})\right),
%\end{equation}
%where we have included the scale factor $a$ for later convenience,
%which we take to be $a\equiv 1$ in flat space-time.

Making use of the decomposition~(\ref{varphi}), where 
 $\varphi(\mathbf{k},t) = \big(2\omega(|\mathbf{k}|)\big)^{-1/2}
                 {\rm e}^{-i\omega(|\mathbf{k}|) t} $,
one finds for the infinite time limit $t\rightarrow \infty$
of the response function~(\ref{response function}) 
\begin{eqnarray}
%\frac{d{\mathcal F}_{\rm flat}(\Delta E)}{dt}  &\!=\!& 
%\int\frac{d^3k}{(2\pi)^3}
% \bigg[
%\frac{\pi\delta\big(\Delta E - \omega(|\mathbf{k}|)\big)}
%         {\omega(|\mathbf{k}|)}
%            \nu(|\mathbf k|)+
%\frac{\pi\delta\big(\Delta E + \omega(|\mathbf{k}|)\big)}
%         {\omega(|\mathbf{k}|)}
%              \left(\nu(|\mathbf k|)+1\right)
% \bigg]
\frac{d{\mathcal F}_{\rm flat}(\Delta E)}{dt}=\frac{k_{\Delta E}}{2\pi}
\left[\nu(k_{\Delta E})\Theta(\Delta E)+
\left(\nu(k_{\Delta E})+1\right)\Theta(-\Delta E)\right]
\,,
\label{response:flat}
\end{eqnarray}
with $k_{\Delta E}\equiv\sqrt{(\Delta E)^2-m^2}$, 
and the $\Theta$-function is defined
by $\Theta(x) = 0$ for $x<0$ and $\Theta(x) = 1$ for $x\geq 0$.
The first term in the square brackets describes particle absorption, induced
by the positive frequency part of the scalar field, the second accounts
for emission, due to the negative frequency contribution.
This result could of course also be
derived by setting $|i\rangle=|0\rangle$, but instead using the Bogolyubov
transformed basis of mode functions
\begin{equation}
\varphi(\mathbf k,t)=
\frac{1}{\sqrt{2\omega(\mathbf k)}}\left(
\alpha(\mathbf k) {\rm e}^{-i\omega(\mathbf k) t}
+\beta(\mathbf k) {\rm e}^{i\omega(\mathbf k) t}
\right)
\label{mode function:fixed number},
\end{equation}
with $\omega(\mathbf k)=\sqrt{\mathbf k^2+m^2}$,
$|\beta(\mathbf k)|^2=\nu(|\mathbf{k}|)$ and 
$|\alpha(\mathbf k)|^2-|\beta(\mathbf k)|^2=1$. When compared to
Eq.~(\ref{response:flat}), an additional term
$2\pi \delta(\Delta E)\int d^3k/(2\pi)^3\,\,
\Re(\alpha \beta^*)/\omega$ arises, which
can be imposed to vanish by choosing the phases of the Bogolyubov coefficients
such that, upon integration, they average to zero.
%More generally, while this contribution may be observable for pure states, 
% it vanishes for a large class of mixed states, upon the appropriate 
%ensemble averaging is performed. 
%One may also regard this contribution as unobservable,
% since it corresponds to zero-energy transitions. 

\section{Detector in Expanding Universes}\label{section:UnruhExpanding}
Due to the fact that formula~(\ref{response:flat}) appropriately describes
a detector which is immersed in a scalar field with given occupation
numbers $\nu(|\mathbf{k}|)$ in flat space, it is tempting to draw the
conclusion that Eq.~(\ref{response function}) will also reproduce the
physical behaviour in the quantum vacuum of an expanding spacetime.
In this section we show that this is however not the case.

At infinitely early times in expanding space-times,
a mode function~(\ref{WKB}) has the asymptotic, negative frequency
flat space-time form~(\ref{mode function:fixed number}) with the trivial
Bogolyubov coefficients $\alpha(\mathbf k)=1$, $\beta(\mathbf k)=0$. When
expanded adiabatically at finite times, the wave function decomposes into an
adiabatically slow varying amplitude and an oscillating phase:
\begin{equation}
\varphi\simeq \frac{1}{(2\omega)^{{1}/{2}}}\left\{
1 +  \frac{1}{4\omega^2}(1-6\xi)\frac{a''}{a}
+\frac{1}{8}\Big(\frac{a''}{a}+\frac{{a'}^2}{a^2}\Big)\frac{a^2 m^2}{\omega^4}
-\frac{5}{16}\frac{{a^\prime}^2}{a^2}\frac{a^4m^4}{\omega^6}
\right\}{\rm e}^{-i\int^\eta W(\mathbf{k},\eta^\prime)d\eta^\prime}
\label{phi:adiabatic}.
\end{equation}
Since the field is still of purely negative frequency, it is clear
that the picture of absorption, induced and spontaneous emission as discussed
in section~\ref{section:UnruhFlat} does not apply here.
 
The canonical Hamiltonian $H[\pi_\phi,\phi,\eta]\equiv H_\phi(\eta)$
has the form of a two-mode squeezed state 
Hamiltonian~\cite{Parker:1969,GarbrechtProkopecSchmidt:2002}
\begin{equation}
H_\phi(\eta)=\frac{1}{2}\int\frac{d^3 k}{2\pi^3} \left\{
\Omega(\mathbf k,\eta)\left(a(\mathbf{k})a^\dagger(\mathbf{k})
+a^\dagger(\mathbf{k}) a(\mathbf{k})\right)
+\left(\Lambda(\mathbf k,\eta) a(\mathbf{k}) a(\mathbf{-k})+h.c.\right)
\right\}\label{Hamiltonian},
\end{equation}
where
\begin{eqnarray}
\Omega\!\!&=&\!\!
\left|\varphi'-\frac{a'}{a}\varphi\right|^2
+\left(\omega^2+6\xi\frac{a''}{a}\right)|\varphi|^2
\,,
\label{Omega}\\
\Lambda\!\!&=&\!\!
\left(\varphi'-\frac{a'}{a}\varphi\right)^2
+\left(\omega^2+6\xi\frac{a''}{a}\right)\varphi^2,
\label{Lambda}
\end{eqnarray}
and $\omega^2(\mathbf k,\eta)=\mathbf{k}^2+a^2(\eta)m^2$. 
For the minimally coupled case, $\xi=0$, a comparison with 
Eq.~(\ref{enden}) reveals that the following identity holds:
\begin{equation}
a^6T^{00}(\mathbf k,\eta)=\frac{1}{2}\Omega(\mathbf k,\eta)
\,,
\label{OmegaT}
\end{equation}
where we defined $\langle 0|T^{00}(x)|0\rangle 
                  = \int [d^3k/(2\pi)^3]T^{00}(\mathbf{k},\eta)$.
We expand~(\ref{Omega}--\ref{Lambda}) 
up to second adiabatic order and find
\begin{eqnarray}
\Omega_{(2)}\!&=&\!
\omega+\frac{1}{2\omega}\left(\frac{a'^2}{a^2}+6\xi\frac{a''}{a}\right)
+\frac{1}{2}\frac{a'^2}{a^2}\frac{a^2m^2}{\omega^3}
+\frac18\frac{a'^2}{a^2}\frac{a^4m^4}{\omega^5}
\,,
\label{Omega:adiabatic}
\\
\Lambda_{(2)}\!&=&\!
 \left\{
\frac{1}{2\omega}\left(\frac{a'^2}{a^2}\!+\!\frac{a''}{a}\right)
+\frac{1}{4}\bigg(
                  \frac{a''}{a}\!+\!3\frac{a'^2}{a^2}
             \bigg)\frac{a^2 m^2}{\omega^3}
-\frac 12 \frac{a'^2}{a^2}\frac{a^4 m^4}{\omega^5}
+i\frac{a'}{a}\bigg(
                    1+\frac12\frac{a^2m^2}{\omega^2}
              \bigg)
 \right\} {\rm e}^{-2i\int^\eta \! \omega d\eta'}
.
\qquad
\label{Lambda:adiabatic}
\end{eqnarray}

When we diagonalize the Hamiltonian~(\ref{Hamiltonian}) by
a Bogolyubov transformation and define
 $\bar\omega^2\equiv \Omega^2-|\Lambda|^2 = \omega^2+6\xi a''/a$,
we get for the particle number~\cite{GarbrechtProkopecSchmidt:2002}
\begin{equation}
n(\mathbf{k},\eta)=
\frac{\Omega(\mathbf k,\eta)}{2\bar\omega(\mathbf k,\eta)}-\frac{1}{2}
,
\label{particle number}
\end{equation}
where the last term corresponds to subtracting the vacuum contribution.
Note that, due to the coupling to the curvature scalar, it may happen that
for certain modes $\bar\omega^2 <0$. This case corresponds to the gravity 
induced spinodal (tachyonic) instability, when particle number is not 
defined. Nevertheless, also in this case $T^{\mu\nu}$ is well defined,
such that it makes sense to study the flow of a suitably regularized
energy density $T^{00}$~\cite{GarbrechtProkopecSchmidt:2002}. 
Upon inserting expression~(\ref{Omega:adiabatic}) into~(\ref{particle number}),
we find that, up to second adiabatic order,
$n(\mathbf k,\eta)$ is independent of the coupling $\xi$ to the curvature,
\begin{equation}
n_{(2)} %\equiv |\beta_{(2)}|^2
         = \bigg[\frac{1}{2\omega}\frac{a^\prime}{a}
                \bigg(1+\frac 12 \frac{a^2m^2}{\omega^2}\bigg)
           \bigg]^2
.
\label{n:2nd order}
\end{equation}
% and can be identified with the
%zeroth order term of the stress-energy~(\ref{ergden}), when taking
%the canonical mode energy to be
%$\varrho^{can}_\mathbf{k}=\Omega_{\mathbf{k}} n_{\mathbf{k}}/2$.
For the minimally coupled case, it follows from Eq.~(\ref{OmegaT})
\begin{equation}
n=\frac{a^6 T^{00}}{\omega}-\frac{1}{2}\label{nT}
\qquad (\xi = 0)
\,,
\end{equation}
which has the very intuitive interpretation that the particle number
is just the total (comoving) energy density divided by the energy
of an individual particle less the cosmological term.
It should not surprise us that this nice interpretation does not
hold when $\xi\not= 0$, because a clear-cut separation of the energy density
into the contribution of the scalar and the gravitational field is not
available.

%We furthermore find,
%that also the second order term matches to the one of the
%stress-energy for the case $\xi=0$, as expected by Eq.~(\ref{OmegaT}).
Note that, despite having only adiabatically slow varying moduli, the
Bogolyubov coefficients $\alpha(\mathbf k,\eta)$ and $\beta(\mathbf k,\eta)$
as used to diagonalize the Hamiltonian~(\ref{Hamiltonian})
are strongly
time-dependent. Since the mode function~(\ref{phi:adiabatic}) is
of purely negative frequency, but the nonvanishing coefficient
$\beta(\mathbf k,\eta)$ yields a positive frequency contribution, this has
to be compensated by oscillating phases.

Let us now consider a detector, whose timescale $\Delta t$, 
during which it is measuring, satisfies
$(\omega/a)^{-1}\ll \Delta t\ll H^{-1}$, that is long enough to
feel the coherence of the scalar field, but much shorter than the Hubble time.
Therefore, we can linearize the differential relation
$dt=ad\eta$ by setting $t=a(\eta_0)\eta$, where $\eta_0$ is a point in time
chosen to be during the period when the measurement is performed.
This means in turn, that the subsequent discussion is only valid
for particle energies in the ultraviolet domain, where
$\omega/a\gg H$.

We introduce $\tilde a=a(\eta_0)$,
$\tilde\omega(\mathbf k)=\omega(\mathbf k,\eta_0)$,
$\tilde\Lambda(\mathbf k,t)
=|\Lambda(\mathbf k,\eta_0)|\exp(-2i\tilde\omega_\mathbf k t/\tilde a)$
and $\tilde\Omega(\mathbf k)=\Omega(\mathbf k,\eta_0)$.
The Hamiltonian~(\ref{Hamiltonian}) can be approximated as the sum of an
unperturbed, time-independent part
\begin{equation}
\tilde H^0_\phi=\frac{1}{2\tilde a}\int\frac{d^3 k}{(2\pi)^3}
\tilde\Omega(\mathbf k) \left( a(\mathbf k)  a^\dagger(\mathbf k)
+  a^\dagger(\mathbf k)  a(\mathbf k)\right),
\end{equation}
%
%where we made use of 
%   $\int d\eta H_\phi \simeq \int d t \tilde H_\phi$
with approximate eigenmodes
\begin{equation}
\tilde\varphi(\mathbf k,t)=\big(2\tilde\omega(\mathbf k)\big)^{-1/2}
{\rm e}^{-i\tilde\omega(\mathbf k)t/\tilde a}\label{phi:eigenmodes},
\end{equation}
and a strongly oscillating perturbation contribution, which accounts for
pair creation and annihilation processes (or alternatively,
two-mode squeezing),
\begin{equation}
\delta \tilde H_\phi(t)=\frac{1}{2\tilde a}\int\frac{d^3 k}{(2\pi)^3} \left(
\tilde \Lambda(\mathbf k,t)  a(\mathbf{k})   a(\mathbf{-k})+h.c.
\right),
\end{equation}
such that $H_\phi\simeq \tilde a\tilde H^0_\phi
                   + \tilde a\delta \tilde H_\phi$.
Note that this type of seperation is not suitable for the case of a given
particle number in flat space as described by the mode function with
time-independent Bogolyubov coefficients~(\ref{mode function:fixed number}),
because then the coefficient of the pair creation and annihilation operator
satisfies $\Lambda=2\alpha\beta$ and is also time-independent.

\begin{figure}[tbp]
\centerline{\hspace{.in} 
\epsfig{file= 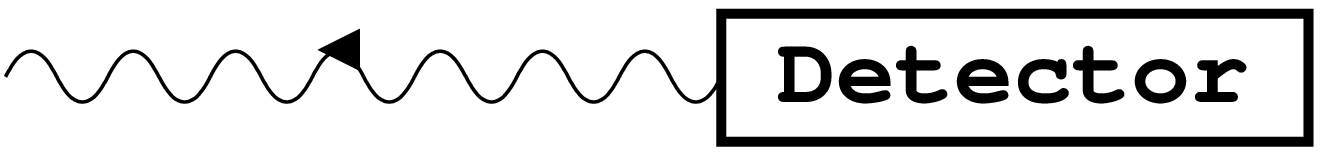, width=3.in,height=0.55in}
\hskip 1cm
\epsfig{file= 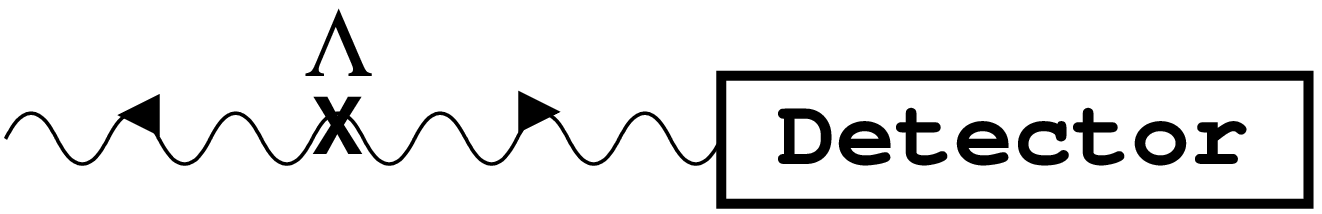, width=3.in,height=0.55in}
}
\vskip -0.1in
\caption{\small
The relevant processes contributing to the response function of an Unruh
detector at order $|\hat h|^2$. The first one corresponds to the 
emission of a particle by the detector, and it is captured by
first order perturbation theory. The second
diagram represents absorption of one of the particles 
of a particle pair created in the expanding background,
and it is captured only at second order in perturbation expansion.
%Only when both processes are taken into account, a consistent answer 
%to the detector response function is obtained. 
}
\label{figure:1}
\end{figure}

With this setup, at second order in time-dependent perturbation theory,
the transition probability of exciting the detector by an energy
$\Delta E$ is given by
\begin{eqnarray}
\mathcal{F}_{(2)}
  \!=\!\int\!\!\frac{d^3k''}{(2\pi)^3}\Bigg|\Big\langle {k''}\Big|
\int_{t_0}^t\!\! dt_1\! \int_{t_0}^{t_1}\! dt_2
&&\!\!\!\!\!\!\int\!\!\frac{d^3k}{(2\pi)^3}
{\rm e}^{i\Delta E t_1}
{\rm e}^{i\mathbf{k}\cdot\mathbf{x}}
\frac{1}{\tilde a}\tilde\varphi(\mathbf k,t_1) a(\mathbf k)
\nonumber\\
\times
&&\!\!\!\!\!\!
\int\!\!\frac{d^3k'}{(2\pi)^3}
      \frac{1}{2\tilde a}\!\tilde\Lambda^*({\mathbf{k}'},t_2)
                  a^\dagger(\mathbf k') a^\dagger(-\mathbf k')
\Big|0\Big\rangle
\Bigg|^2\label{F:integral}.
\end{eqnarray}
For a diagrammatic representation of this process, see the second graph
of Figure~\ref{figure:1}.
Note that it occurs at order $|\hat h|^2$ in the coupling to the
detector, just as the response at first order in perturbation theory.
In Eq.~(\ref{F:integral}), we have not included the
transition amplitude for the creation of three particles.
For this process the time integrations factorize, which means that 
it corresponds to a disconnected graph. Within our approximations,
it does not contribute to the detector response.
There are also processes at order $|\hat h|^2$ 
which contain a higher number of 
$\Lambda$-insertions, which we neglect since they contribute at
higher adiabatic orders.

We evaluate expression~(\ref{F:integral}) using 
Eqs.~(\ref{Lambda:adiabatic}),~(\ref{phi:eigenmodes}),
and take $t\rightarrow\infty$. For the integration, we need
\begin{eqnarray}
\left(2\omega\left(E-\omega\right)\left(E+\omega\right)\right)^2\!\!&&\!\!\!\!\!\!
\left|\int_0^t dt_1 \int_0^{t_1} dt_2 {\rm e}^{i(E-\omega)t_1}
{\rm e}^{2i\omega t_2}\right|^2
\nonumber
\\
=\!\!&&\!\!8\omega^2\!\left[
\sin^2\left(\frac{E+\omega}{2}t\right)
\!+\sin^2\!\left(\frac{E-\omega}{2}t\right)\right]
\nonumber
\\
&-&\!\!8\omega E \left[
\sin^2\!\left(\frac{E+\omega}{2}t\right)
-\sin^2\!\left(\frac{E-\omega}{2}t\right)\right]
+4(E^2-\omega^2)\sin^2\left(\omega t\right)\,,
\end{eqnarray}
and we make use of
\begin{eqnarray}
\lim_{t\to\infty}\frac{\sin^2 \alpha t}{\pi \alpha^2t}=\delta(\alpha)
%\nonumber
.
\end{eqnarray}
The result for the response function is
\begin{equation}
\frac{d{\mathcal F}_{(2)}(\Delta E)}{dt}\simeq 
\frac{k_{\Delta E}}{8\pi} 
               \frac{|\tilde \Lambda|^2}{\tilde \omega^2}
\label{response:2nd order},
\end{equation}
%with
% $\mathbf k$ fixed by the condition
% $\tilde\omega^2/\tilde a^2=(\Delta E)^2$.
where we discarded the contribution from 
the zero mode $\tilde\omega(\mathbf k)=0$ and
where $k_{\Delta E} = \big((\Delta E)^2-m^2\big)^\frac 12$.
Now we employ the relation
$\tilde \Omega^2-| \tilde\Lambda|^2\simeq \tilde\omega^2$
and note that in the ultraviolet domain particle 
numbers~(\ref{particle number}) are small,
 $n \simeq |\tilde \Lambda|^2/(2\tilde\omega)^2 + O(|\tilde \Lambda|^4)\ll 1$,
%and we expand $\tilde\Omega$ in $|\tilde\Lambda|^2/\bar\omega^2$, 
such that we can rewrite~(\ref{response:2nd order}) as
\begin{equation}
\frac{d{\mathcal F}_{(2)}(\Delta E)}{dt}\simeq 
\frac{k_{\Delta E}}{2\pi} n(k_{\Delta E})
\,.
 \label{response:ultraviolet}
\end{equation}
The particle number $n(k_{\Delta E})$ 
can be well approximated by the second adiabatic 
order expression~(\ref{n:2nd order}), which in this context reads
 $n(k_{\Delta E})\simeq
                [H/2\Delta E]^2 [1+m^2/(2\Delta E^2)]^2$, with
$H = a'/a^2$ denoting the physical Hubble parameter. 
Eq.~(\ref{response:ultraviolet}) is our main result
 and establishes a linear relation to the
particle number~(\ref{particle number}) and through Eq.~(\ref{nT}) also
to the energy density of a minimally coupled scalar.

 To compare this with the response function in flat space, we 
note that in the ultraviolet limit, $|\Delta E|\gg H \gg m$,
the response function at first order in perturbative expansion
({\it cf.} the first diagram of Figure~\ref{figure:1})
approaches $d{\cal F}_{(1)}/dt =
                           - \Delta E \Theta(-\Delta E)/(2\pi)$
\cite{GarbrechtProkopec:2004}
and thus is dominating for transitions with $\Delta E <0$. When taken together 
with the second order contribution~(\ref{response:ultraviolet}),
a complete agreement with the flat space response~(\ref{response:flat})  
is reached.

\section{Conclusions}\label{section:Conclusions}

 We have shown that the most important source for the response
of an Unruh detector coupled to a scalar field in an expaning Universe
arises at second adiabatic order. 
 It corresponds to the particle number obtained from
diagonalizing the Hamiltonian, which
agrees for minimal coupling to the curvature with the number density
as inferred from the stress-energy tensor. Thus, 
the latter quantity is observable by a detector, 
which questions the usual approach to stress-energy
renormalization and suggests instead a regularization by a physical cutoff.

The ultraviolet contributions at second adiabatic
order might then cause a considerable back-reaction on the expansion
rate.  The energy density of a minimally coupled massless field
can be easily obtained from~(\ref{T:adiabatic})
\begin{equation}
\varrho_{(2)}
  =\int\limits_0^{a\lambda}\frac{d^3k}{(2\pi)^3}
\frac{1}{4|\mathbf{k}|}\frac{\dot{a}^2}{a^4}
=\frac{\lambda^2}{16\pi^2}H^2,
\end{equation}
where $dt=ad\eta$, $\dot{}\equiv d/dt$, 
 $H=\dot{a}/a$ is the Hubble parameter and $\lambda$ is a physical cutoff. Then
$\varrho_{(2)}=\gamma m_{Pl}^2H^2$ with $\gamma=\lambda^2/(16\pi^2 m_{Pl}^2)$,
where $m_{Pl} \equiv G^{-1/2} \simeq 1.2\times 10^{19}$~GeV
denotes the Planck mass.
The cutoff $\lambda$ should preserve covariant energy conservation
$\dot{\varrho}_{(2)}+3\frac{\dot{a}}{a}(\varrho_{(2)}+p_{(2)})=0$, such that
\begin{equation}
p_{(2)}=-\frac{\gamma m_{Pl}^2}{3}\left(2\frac{\ddot{a}}{a}
+\frac{\dot{a}^2}{a^2}\right)\label{rhoG}.
\end{equation}
The modified Friedmann equation then reads
\begin{equation}
\frac{\ddot{a}}{a}=-\frac{4\pi}{3m_{Pl}^2}\left(
\varrho+3p+\varrho_{(2)}+3p_{(2)}\right)
=-\frac{4\pi}{3m_{Pl}^2}\left(\varrho+3p-2\gamma m_{Pl}^2\frac{\ddot{a}}{a}\right)
\label{FrEff}.
\end{equation}
The cutoff could be due to unknown Planckian effects, in case of which
$\gamma$ should be of order one, or due to a smaller interaction scale.
It is therefore conceivable, that $\gamma$ was varying throughout the
history of the Universe, in particular at structure formation, when particle
densities, which possibly disturb particle production
in the expanding background, dropped drastically in the intergalactic voids.
If today, $\gamma>3/(8\pi)$, gravitation
on cosmological scales might even be repulsive, such that the change 
in $\gamma$ would lead to cosmic acceleration,
providing thus an explanation
for recent observations~\cite{Perlmutter:1998,Riess:1998,Spergel:2003},
without resorting to a cosmological term or an effective field 
theoretical description arising {\it e.g.} from fields like quintessence.

Finally, we point out that our results of
section~\ref{section:UnruhExpanding} apparently
disagree with the notion that the particle spectrum a detector observes in
de Sitter space is thermal. The detector response does not fall exponentially
with the particle energy but rather, just like the energy density, as
a power law,
as indicated by the expressions~(\ref{T:adiabatic})
and~(\ref{Omega:adiabatic}). The analysis here is
restricted to particle momenta larger than the expansion rate of the
Universe. However, it was shown in Ref.~\cite{GarbrechtProkopec:2004}
that at first order in time-dependent perturbation theory, also for
modes comparable to horizon length and larger, the detector response
significantly deviates from a thermal answer.
Therefore, on all momentum scales, an observer in de Sitter space does not see
thermal radiation, a conclusion which is a consequence of the reconciliation of
Unruh's detector response with the stress-energy of a
scalar field in an expanding Universe.


\begin{thebibliography}{99}
\bibliographystyle{unsrt}

\bibitem{GibbonsHawking:1977}
G.~W.~Gibbons and S.~W.~Hawking,
``Cosmological Event Horizons, Thermodynamics, And Particle Creation,''
Phys.\ Rev.\ D {\bf 15} (1977) 2738.
%%CITATION = PHRVA,D15,2738;%%

\bibitem{Lapedes:1978}
A.~S.~Lapedes,
``Bogolyubov Transformations, Propagators, And The Hawking Effect,''
J.\ Math.\ Phys.\  {\bf 19} (1978) 2289.
%%CITATION = JMAPA,19,2289;%%

\bibitem{Higuchi:1986}
A.~Higuchi,
``Quantization Of Scalar And Vector Fields Inside The Cosmological Event
Horizon And Its Application To Hawking Effect,''
Class.\ Quant.\ Grav.\  {\bf 4} (1987) 721.
%%CITATION = CQGRD,4,721;%%

\bibitem{BoussoMaloneyStrominger:2001}
R.~Bousso, A.~Maloney and A.~Strominger,
``Conformal vacua and entropy in de Sitter space,''
Phys.\ Rev.\ D {\bf 65} (2002) 104039
[arXiv:hep-th/0112218].
%%CITATION = HEP-TH 0112218;%%

\bibitem{GarbrechtProkopec:2004}
B.~Garbrecht and T.~Prokopec,
``Unruh response functions for scalar fields in de Sitter space,''
arXiv:gr-qc/0404058.
%%CITATION = GR-QC 0404058;%%

\bibitem{BirrellDavies:1982}
N.~D.~Birrell and P.~C.~W.~Davies,
``Quantum Fields In Curved Space,''
Cambridge, UK: Univ. Pr. (1982).

\bibitem{MukhanovChibisov:1981}
V.~F.~Mukhanov and G.~V.~Chibisov,
``Quantum Fluctuation And 'Nonsingular' Universe. (In Russian),''
JETP Lett.\  {\bf 33} (1981) 532
[Pisma Zh.\ Eksp.\ Teor.\ Fiz.\  {\bf 33} (1981) 549].
%%CITATION = JTPLA,33,532;%%

\bibitem{GuthPi:1982}
A.~H.~Guth and S.~Y.~Pi,
``Fluctuations In The New Inflationary Universe,''
Phys.\ Rev.\ Lett.\  {\bf 49} (1982) 1110.
%%CITATION = PRLTA,49,1110;%%

\bibitem{Hawking:1982}
S.~W.~Hawking,
``The Development Of Irregularities In A Single Bubble Inflationary Universe,''
Phys.\ Lett.\ B {\bf 115} (1982) 295.
%%CITATION = PHLTA,B115,295;%%

\bibitem{Starobinsky:1982}
A.~A.~Starobinsky,
``Dynamics Of Phase Transition In The New Inflationary Universe Scenario And
Generation Of Perturbations,''
Phys.\ Lett.\ B {\bf 117} (1982) 175.
%%CITATION = PHLTA,B117,175;%%

\bibitem{Parker:1969}
L.~Parker,
``Quantized Fields And Particle Creation In Expanding Universes. 1,''
Phys.\ Rev.\  {\bf 183} (1969) 1057.
%%CITATION = PHRVA,183,1057;%%

\bibitem{Hawking:1974}
S.~W.~Hawking,
``Particle Creation By Black Holes,''
Commun.\ Math.\ Phys.\  {\bf 43} (1975) 199.
%%CITATION = CMPHA,43,199;%%

\bibitem{Unruh:1976}
W.~G.~Unruh,
``Notes On Black Hole Evaporation,''
Phys.\ Rev.\ D {\bf 14} (1976) 870.
%%CITATION = PHRVA,D14,870;%%

\bibitem{ParkerFulling:1974}
L.~Parker and S.~A.~Fulling,
``Adiabatic Regularization Of The Energy Momentum Tensor Of A Quantized Field
In Homogeneous Spaces,''
Phys.\ Rev.\ D {\bf 9} (1974) 341.
%%CITATION = PHRVA,D9,341;%%

\bibitem{HabibMolina-ParisMottola:1999}
S.~Habib, C.~Molina-Paris and E.~Mottola,
``Energy-momentum tensor of particles created in an expanding universe,''
Phys.\ Rev.\ D {\bf 61} (2000) 024010
[arXiv:gr-qc/9906120].
%%CITATION = GR-QC 9906120;%%

%\cite{Chernikov:zm}
\bibitem{ChernikovTagirov:1968}
N.~A.~Chernikov and E.~A.~Tagirov,
``Quantum Theory Of Scalar Fields In De Sitter Space-Time,''
Annales Poincare Phys.\ Theor.\ A {\bf 9} (1968) 109.
%%CITATION = AHPAA,A9,109;%%

\bibitem{BunchDavies:1978}
T.~S.~Bunch and P.~C.~Davies,
``Quantum Field Theory In De Sitter Space: 
 Renormalization By Point Splitting,''
Proc.\ Roy.\ Soc.\ Lond.\ A {\bf 360} (1978) 117.
%%CITATION = PRSLA,A360,117;%%

\bibitem{Christensen:1976}
S.~M.~Christensen,
``Vacuum Expectation Value Of The Stress Tensor In An Arbitrary Curved
Background: The Covariant Point Separation Method,''
Phys.\ Rev.\ D {\bf 14} (1976) 2490.
%%CITATION = PHRVA,D14,2490;%%

\bibitem{'tHooftVeltman:1974}
G.~'t Hooft and M.~J.~G.~Veltman,
``One Loop Divergencies In The Theory Of Gravitation,''
Annales Poincare Phys.\ Theor.\ A {\bf 20} (1974) 69.
%%CITATION = AHPAA,A20,69;%%

\bibitem{Donoghue:1994}
J.~F.~Donoghue,
``General Relativity As An Effective Field Theory: The Leading Quantum
Corrections,''
Phys.\ Rev.\ D {\bf 50} (1994) 3874
[arXiv:gr-qc/9405057].
%%CITATION = GR-QC 9405057;%%

%\cite{Bjerrum-Bohr:2002ks}
\bibitem{Bjerrum-Bohr:2002ks}
N.~E.~J.~Bjerrum-Bohr, J.~F.~Donoghue and B.~R.~Holstein,
``Quantum corrections to the Schwarzschild and Kerr metrics,''
Phys.\ Rev.\ D {\bf 68} (2003) 084005
[arXiv:hep-th/0211071].
%%CITATION = HEP-TH 0211071;%%

%\cite{Bjerrum-Bohr:2002kt}
\bibitem{Bjerrum-Bohr:2002kt}
N.~E.~J.~Bjerrum-Bohr, J.~F.~Donoghue and B.~R.~Holstein,
``Quantum gravitational corrections to the nonrelativistic scattering potential
of two masses,''
Phys.\ Rev.\ D {\bf 67} (2003) 084033
[arXiv:hep-th/0211072].
%%CITATION = HEP-TH 0211072;%%

%\cite{DonoghueHolsteinGarbrechtKonstandin:2001}
\bibitem{DonoghueHolsteinGarbrechtKonstandin:2001}
J.~F.~Donoghue, B.~R.~Holstein, B.~Garbrecht and T.~Konstandin,
``Quantum corrections to the Reissner-Nordstroem and Kerr-Newman  metrics,''
Phys.\ Lett.\ B {\bf 529} (2002) 132
[arXiv:hep-th/0112237].
%%CITATION = HEP-TH 0112237;%%

%\cite{Bjerrum-Bohr:2002sx}
\bibitem{Bjerrum-Bohr:2002sx}
N.~E.~J.~Bjerrum-Bohr,
``Leading quantum gravitational corrections to scalar QED,''
Phys.\ Rev.\ D {\bf 66} (2002) 084023
[arXiv:hep-th/0206236].
%%CITATION = HEP-TH 0206236;%%

%\cite{Kluger:1991ib}
\bibitem{Kluger:1991ib}
Y.~Kluger, J.~M.~Eisenberg, B.~Svetitsky, F.~Cooper and E.~Mottola,
``Pair production in a strong electric field,''
Phys.\ Rev.\ Lett.\  {\bf 67} (1991) 2427.
%%CITATION = PRLTA,67,2427;%%

%\cite{Kluger:1992gb}
\bibitem{Kluger:1992gb}
Y.~Kluger, J.~M.~Eisenberg, B.~Svetitsky, F.~Cooper and E.~Mottola,
``Fermion pair production in a strong electric field,''
Phys.\ Rev.\ D {\bf 45} (1992) 4659.
%%CITATION = PHRVA,D45,4659;%%

%\cite{GarbrechtProkopecSchmidt:2002}
\bibitem{GarbrechtProkopecSchmidt:2002}
B.~Garbrecht, T.~Prokopec and M.~G.~Schmidt,
``Particle number in kinetic theory,''
arXiv:hep-th/0211219.
%%CITATION = HEP-TH 0211219;%%

\bibitem{Perlmutter:1998}
S.~Perlmutter {\it et al.}  [Supernova Cosmology Project Collaboration],
``Measurements of Omega and Lambda from 42 High-Redshift Supernovae,''
Astrophys.\ J.\  {\bf 517} (1999) 565
[arXiv:astro-ph/9812133].
%%CITATION = ASTRO-PH 9812133;%%

\bibitem{Riess:1998}
A.~G.~Riess {\it et al.}  [Supernova Search Team Collaboration],
``Observational Evidence from Supernovae for an Accelerating Universe and a
Cosmological Constant,''
Astron.\ J.\  {\bf 116} (1998) 1009
[arXiv:astro-ph/9805201].
%%CITATION = ASTRO-PH 9805201;%%

\bibitem{Spergel:2003}
D.~N.~Spergel {\it et al.},
``First Year Wilkinson Microwave Anisotropy Probe (WMAP) Observations:
Determination of Cosmological Parameters,''
Astrophys.\ J.\ Suppl.\  {\bf 148} (2003) 175
[arXiv:astro-ph/0302209].
%%CITATION = ASTRO-PH 0302209;%%

\end{thebibliography}
\end{document}